\documentstyle[aps,multicol]{revtex}

\renewcommand{\narrowtext}{\begin{multicols}{2} \global\columnwidth20.5pc}
\renewcommand{\widetext}{\end{multicols} \global\columnwidth42.5pc}
\newcommand{\p}{\partial}
\newcommand{\Tr}{{\rm Tr}}
\newcommand{\Ad}{{\rm Ad}}
\newcommand{\diag}{{\rm diag}}

\begin{document}

\title{$D$-branes in the WZW model}

\author{Anton Yu. Alekseev$^*$,
 Volker Schomerus$^{\dagger}$}

\address{$^*$
Institute for Theoretical Physics, Uppsala University,
Box 803, S-75108, Uppsala, Sweden}

\address{ $^\dagger$
II. Institut f\"ur Theoretische Physik, Universit\"at 
Hamburg, Luruper Chaussee 149, D-22761 Hamburg, Germany}

\date{November 1998}

\maketitle

{\tightenlines
\begin{abstract}
It is stated in the literature that $D$-branes in the WZW-model
associated with the gluing condition $J= - \bar{J}$ along the boundary 
correspond to branes filling out the whole group volume. We show 
instead that the end-points of open strings are rather bound to 
stay on `integer' conjugacy classes. In the case of $SU(2)$ level 
$k$ WZW model we obtain $k-1$ two dimensional Euclidean $D$-branes
and two $D$ particles sitting at the points $e$ and $-e$.
\vskip 0.1cm


\end{abstract}
}

\narrowtext

String theory on a group manifold can be described by
the world-sheet Wess-Zumino-Witten (WZW) action,
\begin{equation} \label{WZW}
S(g) = \frac{k}{8\pi} \int \Tr \left( (\p_\mu g g^{-1})^2
+ \frac{2}{3} d^{-1} (dg g^{-1})^3 \right) . 
\end{equation}
This theory possesses chiral currents ( with $\p_\pm = 
\p_t \pm \p_x $), 
\begin{equation} \label{glue}
J= - \p_+ g g^{-1} \ , \ \bar{J}=g^{-1} \p_- g,
\end{equation}
Let us perform our analysis of branes in the closed string 
picture where $D$-branes are described as special `initial 
conditions' for closed strings rather than by boundary 
conditions in a theory of open strings. We consider 
$D$-branes corresponding to the standard gluing condition 
$J= - \bar{J}$ at the initial time $t=t_0$. The same gluing 
condition was used in \cite{Ishi}. For comparison with gluing 
conditions in the open string picture one needs to include 
an extra factor $-1$ coming from the 
transformation properties of currents under coordinate 
transformations of the world-sheet \cite{Cardy}. 

$D$-branes of this type were studied in the literature. 
For instance, Kato and Okada \cite{Kato} suggest that they 
correspond to Neumann boundary conditions in all directions 
and, hence, that they fill the whole group manifold $G$. The 
same assertion is implicitly contained in \cite{Ishi} where 
the gluing condition $J= - \bar{J}$ is considered as a 
generalization of Neumann boundary conditions for a free 
bosonic string. This is clearly not the case: If we insert
the parametrization $g = \exp (X)$ of the group valued 
field $g$ near the group unit into the gluing conditions 
we obtain $\partial_x X = 0$, i.e.\ the derivative of 
$X$ along the boundary vanishes. Hence, one should rather 
view the relation $J = -\bar{J}$ as a  generalization of 
Dirichlet boundary conditions along the boundary. Using 
this argument, Stanciu and Tseytlin \cite{Stanciu} 
(in the context of Nappi-Witten backgrounds) see a rather 
point-like structure of the associated $D$-branes. 
Our findings fit well with the analysis of Klimcik
and Severa \cite{Klimcik}: they identify $D$-branes in the WZW
model with orbits of dressing transformations.
If the `double' (used in \cite{Klimcik}) is chosen as 
$G \times G$, the dressing orbits coincide
with conjugacy classes (see \cite{AMM} for details).
Note, however, that no gluing conditions
are specified in \cite{Klimcik}.
 
The analysis below will show that the end-points of open strings 
with gluing conditions $J = - \bar{J}$ (in the closed string 
picture) are localized on special `integer' conjugacy classes 
$ghg^{-1}$ for some fixed $h$. In particular, for the $SU(2)$ 
level $k$ WZW model we obtain two $D$-particles at the points 
$\pm e$  and $k-1$ two dimensional Euclidean $D$-branes.

In terms of $\p_t,\p_x$, the gluing condition $J=-\bar J$ reads
\begin{equation} \label{reads}
g^{-1} \p_t g - \p_t g g^{-1} = g^{-1} \p_x g + \p_x g g^{-1}.
\end{equation}
It is convenient to introduce a special notation for the adjoint
action of $G$ on its Lie algebra, $\Ad(g) y = g y g^{-1}$. Then, 
equation (\ref{reads}) can be rewritten as
\begin{equation} \label{Ad}
(1- \Ad(g)) \, g^{-1} \p_t g = (1 + \Ad(g) )\, g^{-1} \p_x g.
\end{equation}
We split the tangent space to the group $G$ at the point $g$
into an orthogonal (with respect to the Killing metric)
sum, $T_gG= T_g^\perp G \oplus T_g^{\|} G$, where $ T_g^{\|} G$
consists of vectors tangential to the orbit of $\Ad$ through $g$.
Observe that on $T_g^\perp G$ the operator $1- \Ad(g)$ vanishes
whereas $1+ \Ad(g) =2$. Hence, we conclude that 
\begin{equation}
(g^{-1} \p_x g)^{\perp} =0,
\end{equation}
and the corresponding $D$-branes coincide with the conjugacy classes.
In the open string picture (which has $t$ and $x$ exchanged), the 
previous equation is a Dirichlet-type condition for components 
orthogonal to the conjugacy class. 

If we restrict our consideration to some
conjugacy class $C$, the operator $(1-\Ad(g))$ acting on the
tangent space $T_g^{||}G$ becomes invertible,
and equation (\ref{Ad}) can be rewritten in the form
\begin{equation}
g^{-1} \p_t g= \frac{1 + \Ad(g)}{1- \Ad(g)} g^{-1} \p_x g.
\end{equation}
Thus, it gives rise to a 2-form ($B$-field) on the conjugacy class
\begin{equation}\label{7} 
\omega= \frac{k}{8\pi} \Tr \left( 
g^{-1} dg \frac{1 + \Ad(g)}{1- \Ad(g)} g^{-1} dg \right),
\end{equation}
where we have taken into account the normalization
of the action (\ref{WZW}).
The 2-form $\omega$ is not closed, instead
\begin{equation} \label{8} 
d\omega = -\frac{k}{12\pi} \Tr (dg g^{-1})^3.
\end{equation}
According to \cite{Klimcik}, $D$-branes in the WZW model are
specified by a choice of a submanifold $D \subset G$
such that the restriction of the Wess-Zumino form
$\eta=-\frac{k}{12\pi} \Tr (dg g^{-1})^3$ to $D$ is exact,
together with a 2-form $\omega$ on $D$ such that
$d \omega = \eta$. Equation (\ref{8}) shows that 
conjugacy classes satisfy this condition, and that
the form (\ref{7}) gives a canonical choice of the primitive
$\omega$. Conjugacy classes equipped with such 2-forms were
considered in  \cite{AMM} as examples of Hamiltonian
spaces which admit  group-valued moment maps.

Together, the $B$-field $\omega$ and the 
topological Wess-Zumino term in (\ref{WZW}) impose
a further constraint on the choice of conjugacy classes
which can be used as $D$-branes. For simplicity, we analyze
it only in the case of $G=SU(2)$. A $D$-brane in the target
space corresponds to a boundary state of the world-sheet
theory. In the case of WZW model such a boundary state
can be visualized as wave functional $\Psi(g(x))$ on the
space of closed loops $g(x)$ in some conjugacy class $C$.
Typical conjugacy classes in $G=SU(2)$ (other than $e$ and $-e$)
are 2-spheres. So, a closed loop on $C$
can be contracted in two different
ways giving rise to an ambiguity in the phase of the wave
functional
\begin{equation} \label{quantization}
\Delta \phi = \int_{C} \omega+ \frac{k}{12\pi} \int_{B} \Tr (dg g^{-1})^3,
\end{equation}
where $B$ is one of the 3-balls in $SU(2)=S^3$ bounded by
the conjugacy class $C$.  
Boundary states correspond to conjugacy
classes with $\Delta \phi = 2\pi j$ with integer $j$. 
Equation
(\ref{quantization}) generalizes the standard Bohr-Sommerfeld
quantization condition $\int \omega = 2\pi j$.
An elementary calculation (see
\cite{ASS}, equation (C.21)) shows that the conjugacy classes 
corresponding to $\Delta \phi = 2\pi j, j=1, \dots, k-1$ pass 
through the points $\diag(\exp(\pi i j/k), \exp(-\pi i j/k))$. 
The point-like conjugacy classes $e$ and $-e$ allow for an 
unambiguous choice of the wave functional $\Psi(g)=1$ and 
correspond to $D$-particles.

These findings give a complete list of boundary conditions 
associated with the gluing condition $J = - \bar J$ which 
is in perfect agreement with the results of Cardy \cite{Cardy}. 
Other boundary conditions can come only from different choices 
of the gluing condition. In the simplest case we modify the  
relation $J = - \bar J$ by acting with an inner automorphism 
of the finite Lie algebra on the left hand side. In contrast 
to what was suggested in \cite{Kato}, such a shift with inner 
automorphisms cannot possibly change the geometry of branes. It
is rather associated with symmetries of the target, namely with 
the right action of $G$ on itself (see e.g. \cite{ReSc}). In 
case of $G = SU(2)$, our discussion exhausts all boundary 
conditions with a maximal chiral Kac-Moody symmetry. Let us
note that the non-abelian structure of the Kac-Moody algebra 
places severe constraints for the construction of boundary 
conditions which preserve a non-abelian symmetry (see e.g. 
\cite{Kato}). In particular, it forbids the reversal of signs 
in all directions of the Lie algebra (i.e. the gluing condition 
$J = \bar{J}$) which is used in abelian theories to manufacture 
e.g. volume filling branes from D-particles etc.

If the Lie algebra of $G$ admits outer automorphisms there 
can be additional boundary conditions with maximal symmetry 
which are not covered by our analysis above. The same applies 
to boundary conditions that do not preserve the full chiral 
Kac-Moody symmetry of the WZW model. The construction and
interpretation of such branes remains an interesting open 
problem. 
  
\begin{acknowledgements} 
We thank J. Fuchs, C. Klimcik, A. Recknagel, C. Schweigert, 
S. Stanciu, A. Tseytlin and T. Strobl for their useful comments and 
critical remarks. The stay of V.S. at the ITP in Uppsala 
was supported by the DAAD.  
\end{acknowledgements}

\widetext

\end{document}